\documentclass[conference]{IEEEtran}
\IEEEoverridecommandlockouts
\usepackage{cite}
\usepackage{amsmath,amssymb,amsfonts}
\usepackage{graphicx}
\usepackage{textcomp}
\usepackage{xcolor}
\usepackage{algorithm, algpseudocode}
\usepackage{soul} % for the command \hl
\usepackage{array}
\usepackage{siunitx}

\newcommand{\sysname}[0]{\textit{SEASONS}}

\def\BibTeX{{\rm B\kern-.05em{\sc i\kern-.025em b}\kern-.08em
    T\kern-.1667em\lower.7ex\hbox{E}\kern-.125emX}}
\begin{document}

\title{SEASONS: Signal and Energy Aware Sensing on iNtermittent Systems }
\author{
	\IEEEauthorblockN{1\textsuperscript{st} Pouya Mahdi Gholami}
	\IEEEauthorblockA{ \textit{University of Chicago}}\\
\and
	\IEEEauthorblockN{2\textsuperscript{nd} Henry Hoffmann}
	\IEEEauthorblockA{ \textit{University of Chicago}}
}

\maketitle
\begin{abstract}
Both energy-aware, batteryless intermittent systems and signal-aware adaptive sampling algorithms (ASA) aim to maximize sensor data accuracy under energy constraints in edge devices. Intuitively, combining both into a signal- \& energy-aware solution would yield even better accuracy. Unfortunately, ASAs and intermittent systems rely on conflicting energy availability assumptions. So, a straightforward combination cannot achieve their combined benefits.

Therefore, we propose \sysname{}, the first framework for signal- and energy-aware intermittent systems. \sysname{} buffers signal data in time, monitoring queue dynamics to ensure the data is reported within a user-specified latency constraint. \sysname{} improves sensor data accuracy by 31\% without increasing energy.

\end{abstract}

\begin{IEEEkeywords}
self-aware, intermittent, adaptive sampling
\end{IEEEkeywords}

\section{Introduction}\label{sec:intro}

\textit{Batteryless intermittent} sensing systems are a recent advancement in embedded edge sensing \cite{Maeng_CatNap, Lucia_Dino, Desai_Camaroptera, Colin_Chain, Gobieski_intelligence, Hester_MayFly,Yildrim_InK, Denby_Orbital, CircusMaximus, Bakar_REHASH, Balsamo_Hibernus}. Intermittent systems sample environmental signals to provide accurate sensor data to a central server while using only harvested ambient energy stored in small capacitors. Batteryless operation removes the many disadvantages of batteries (\textit{e.g.} battery replacement, system size, and environmental concerns), making intermittent systems suitable for many domains from small satellites \cite{Denby_Orbital} to infrastructure monitoring \cite{CircusMaximus}. However, accurate sensing is challenging given the unreliability of ambient energy sources such as solar. Thus, \textit{energy-awareness}---\textit{i.e.} adapting to scarce and variable energy---is an essential for intermittent runtime systems. 

\textit{Signal-awareness}, achieved via a family of Adaptive Sampling Algorithms (ASA), is another advancement in the field of sensing systems \cite{ASA_Deviation, ASA_Linear, ASA1, ASA_Reinf1, ASA_SkipRNN}. ASAs balance the trade-off between energy and accuracy in an optimal manner by analyzing the sensed environmental signals. They ensure that energy is used to collect \textit{significant samples}---i.e., those that have the most influence on accuracy---and they avoid spending energy on less significant samples. 
Compared to non-adaptive techniques, ASAs improve the accuracy of sensed data and reduce the system energy requirements. Overall, both intermittent systems and ASAs aim to provide accurate sensor data 
despite energy limitations. Intuitively, these approaches should be combined into a signal- and energy-aware intermittent system to achieve the advantages of both. 

Unfortunately, despite their shared goals, these two approaches rely on incompatible assumptions. ASAs achieve their goal through sampling rate variation: significant sensor data are sampled at higher frequency, whereas less significant ones are sampled less. Since the sampling rate directly affects energy consumption, varying sampling rates produces fluctuating energy consumption. Windows of significant samples consume more energy and the less significant ones do the opposite. ASAs assume that these periods of high and low energy draw will balance over a battery’s lifetime (lasting potentially hours or days). However, intermittent sensors use capacitors that only store energy for milliseconds of operation \cite{Lucia_Dino}. Thus, they cannot sustain long periods of high energy consumption nor can they store the extra energy from low energy consumption windows. In short, existing intermittent systems cannot support the variable energy consumption required by ASAs.

To address this problem, we introduce \sysname{}, a framework that enables signal-aware ASA operation on energy-aware intermittent systems. An ASA operating via \sysname{} schedules and budgets to achieve a constant energy consumption, regardless of the signal. To achieve this, \sysname{} uses time buffering. When the ASA needs to increase the sampling rate (because it has encountered significant samples), \sysname{} collects the data, and immediately enqueues it, delaying processing and communication to maintain constant energy consumption. Doing so mitigates a period of high energy consumption typical of an ASA. When the ASA reduces the sampling rate, \sysname{} fast tracks the processing and communication of the enqueued samples, maintaining the constant rate of energy consumption. The fast tracking uses up the the extra energy that could not be stored over a long term, avoiding wastage of surplus energy.

While time buffering resolves the incompatibility between the two approaches, it can cause arbitrarily delays in communicating the sensed data. Expired samples, data that has not been sent within application defined latency constraints \cite{Subatovich_Ocelot, Hester_MayFly}, occur more frequently with time buffering. \sysname{} avoids such scenarios by analyzing queue dynamics to adjust sampling rates dynamically and ensure all collected data is processed and communicated within the latency constraint.

With \sysname{}, energy-aware intermittent systems can employ signal-aware sampling policies to reduce the impact of low energy operation on sensor data accuracy. We find \sysname{} improves data accuracy by an average of 31\% across multiple sensor datasets and energy budgets compared to prior purely energy-aware intermittent systems.

\begin{figure*}
    \centering
    \includegraphics[width=\textwidth]{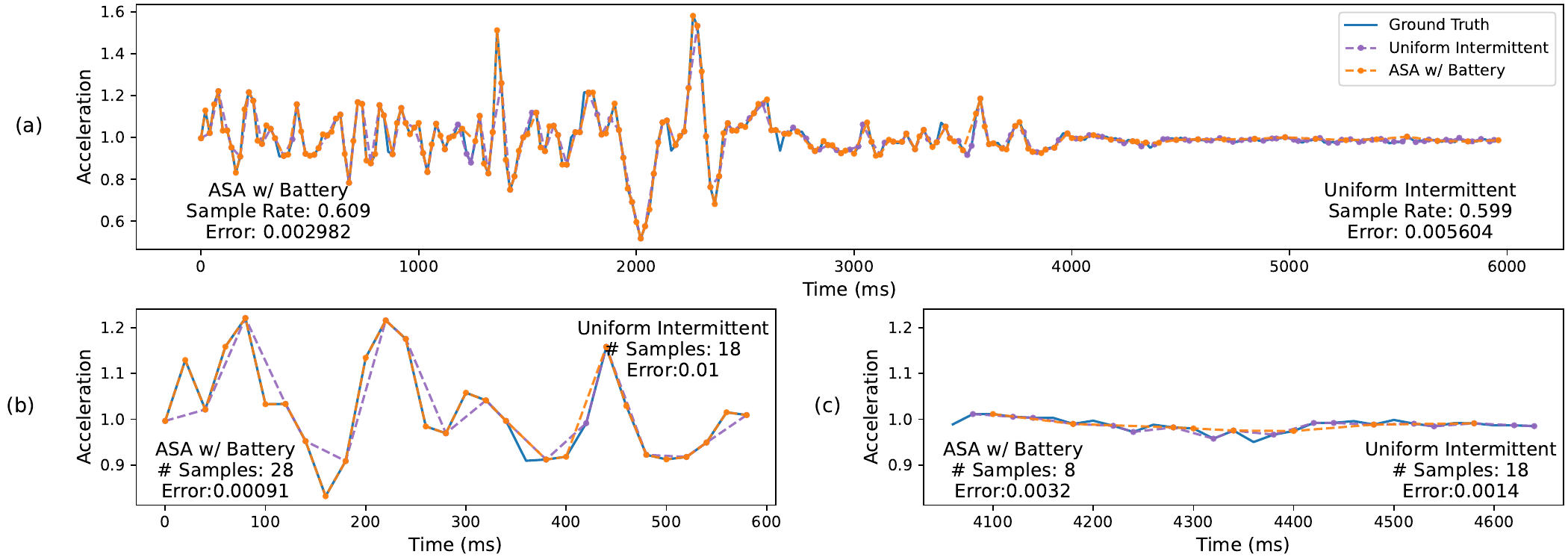}
    \caption{(a): Energy-Aware intermittent uniform sampling (purple) vs battery-backed ASA (orange) of accelerometer data over 6 seconds w/ 60\% collection rate. The blue line (labeled ground truth) represents the real accelerometer signal sampled at 50Hz. (b): Behavior of sampling policies over a highly volatile phase. Adaptive sampling collects 10 more samples and achieves one order of magnitude lower error than uniform sampling. Since the adaptive method takes more samples, it consumes more energy than the uniform over this period. (c): Behavior of subsampling policies over a low volatility phase. The uniform strategy collects 10 more samples and achieves 2.2x lower error. Since the adaptive method takes less samples, it consumes less energy than uniform.}
    \label{fig:battery-based}
\end{figure*}

This work makes the following contributions:
\begin{itemize}
    \item We analyze how and why straightforward implementation of ASAs on intermittent systems fail. 
    \item We propose practical solutions that circumvent these failures and present \sysname{}, a novel framework that enables ASA operation on intermittent systems.
\end{itemize}

To our knowledge, we are the first to study adaptive sampling on intermittent systems, formalize the shortcomings of naive adoption, and offer a solution.  We provide the first signal \& energy-aware intermittent interface. Furthermore, we are the first to point out the importance of queue dynamics on an intermittent system and capitalize on it. 

\section{Background \& Motivation}\label{sec:background}
We consider a common low-power embedded sensing device with the goal of providing accurate and timely data to a central server. This design is common in traditional embedded sensing \cite{ZebraNet} or even emerging intermittent setups \cite{CircusMaximus}. Purely Energy-aware Intermittent designs achieve this goal by altering sensor data quality based on available energy. A popular and common approach is uniformly reducing sample collection rate  \cite{Bakar_REHASH, Maeng_CatNap}. In contrast,  battery-backed signal- \& energy-aware systems under the same energy constraints use an ASA to improve data accuracy by devoting energy to the most significant data samples \cite{ASA_Deviation, ASA_Linear, ASA1, ASA_Reinf1, ASA_SkipRNN}.

Figure \ref{fig:battery-based} shows a comparison between the two approaches. The blue line shows an  accelerometer signal sampled at 50Hz. The purple and orange lines show the inferred signal from the intermittent system and the ASA, respectively. Both the intermittent and battery-backed system are under energy constraints that permit collection of just 60\% of possible samples; however,  the ASA strategy results in 88\% improvement in data accuracy compared to the intermittent result.

The improvement is due to the ASA's selective collection of the samples. Unlike the intermittent system, the ASA equipped battery-backed system does not uniformly collect 6 samples out of 10 possible samples. Since the accelerometer signal begins with highly volatile behavior (figure \ref{fig:battery-based} (b))---which the ASA considers significant---the ASA collects almost every possible sample during that time. Later (figure \ref{fig:battery-based} (c)), the ASA considers the less variable signal less significant, so it collects 2 or 3 samples out of 10 possible samples. Over the entire operation window, the ASA improves accuracy under the same energy constraints. Combining this higher accuracy with the batteryless benefits of intermittent systems would be advantageous. However, a straightforward combination of both fails to achieve this higher accuracy.

Typical operation of an ASA assumes a large energy buffer to support variable energy consumption rates. In figure \ref{fig:battery-based} (a), the ASA strategy requires taking significantly more samples in the 0--3000 ms time range compared to the 4000--6000 ms range. The initial increase in sampling rate corresponds to a higher draw rate from the system's energy buffer. Later, the reduced sampling rate reduces energy draw, balancing the total consumed energy across the entire operation window. An intermittent system with a buffer that an only support milliseconds of operation is unable to support the seconds of distance between this energy balancing. Ported directly to an intermittent system, the ASA cannot reliably alter the sampling rate without changing the energy consumption rate and causes power failures, wasting energy and reducing accuracy.

\begin{figure*}
    \centering
    \includegraphics[width=\textwidth]{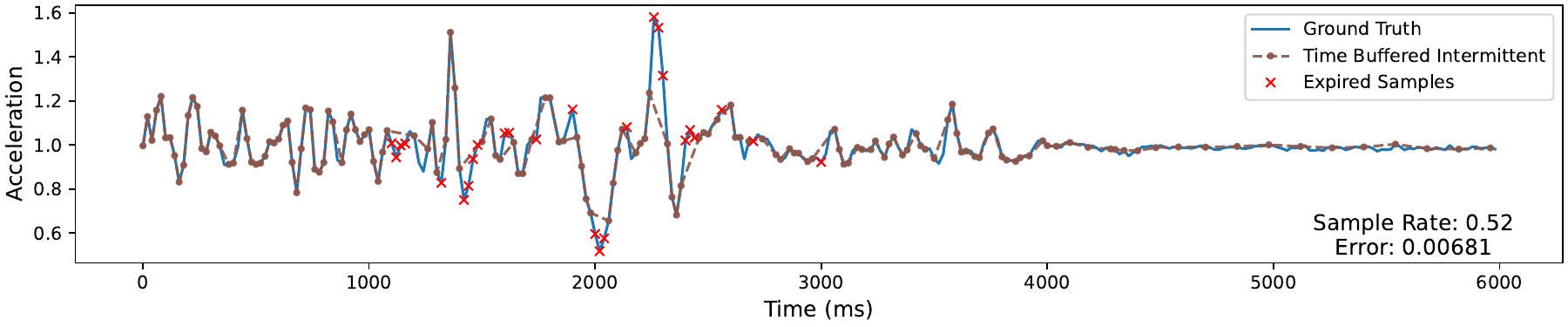}
    \caption{Adaptive subsampling with time buffering and freshness enforcement by dropping samples.}
    \label{fig:interm-adaptive-drops}
\end{figure*}

\section{\sysname{}}\label{sec:operating}
\sysname{} is a novel framework that combines the benefits of batteryless intermittent systems with signal-aware ASAs. \sysname{} resolves the conflicting energy availability assumptions of these system by buffering in time, a concept that relies on the postponability of processing and communication tasks. Time buffering alone, however, is unaware of latency constraints and can lead to latency violations. Such violations reduce the accuracy, so \sysname{} utilizes sample queue dynamics to anticipate and avoid such violations.  This section explains these two key concepts.

\subsection{Adaptive Sampling via Time Buffering}
As studied by prior intermittent work, many parts of an intermittent application do not have a strict deadline \cite{Maeng_CatNap}. Samples are numerous and their deadlines are hard, but processing and communication occur less regularly and their deadlines are more flexible. Thus, when feasible, a system can postpone processing or communication. We dub this operation as \textbf {time buffering}. Further, with time buffering, samples must wait in a non-volatile queue for processing and communication \cite{Hester_MayFly, Maeng_CatNap, Yildrim_InK}. We call this queue the \textit{sample queue}. 

With time buffering, an intermittent system can employ an ASA without altering the energy consumption as follows. Given significant samples, the system increases sampling rate and enqueues samples until later processing (\textit{e.g.} 0--3000 ms time range in figure \ref{fig:battery-based}); given less significant samples, the system reduces sampling rate, fast-tracks sample processing and communication to empty the sample queue (\textit{e.g.} 4000--6000 ms time range in figure \ref{fig:battery-based}).

However, time buffering maintains no control over the number of samples in the queue. In general, the system cannot predict when a less significant window of samples will arrive to balance  the earlier window of significant samples. This unpredictably could cause the latency or queue size to grow arbitrarily long. This is unacceptable for embedded edge devices as they are resource (including memory) limited and must operate within an application-defined latency.

Respecting application latency constraints is vital in modern intermittent runtimes \cite{Hester_MayFly, Subatovich_Ocelot}. This concept, dubbed \textit{freshness}, ensures that samples are processed in a timely manner, despite potential power failures and energy shortages. Once a freshness violation has been detected (\textit{i.e.} a sample has \textit{expired} its latency constraint), these systems drop the sample and prevent it from progressing in the application pipeline so that additional energy is not wasted on processing and communicating an expired sample. 

Figure \ref{fig:interm-adaptive-drops} depicts the operation of a time buffered intermittent system running and ASA with freshness enforcement under the same constraints as figure \ref{fig:battery-based}. While freshness enforcement is vital to prevent energy wastage from sending expired samples, it results in many samples being dropped before they are communicated to the server.  These sample drops then reduce the overall accuracy. In figure \ref{fig:interm-adaptive-drops}, 13\% of taken samples expire before they can be sent, leading to a signal inference that misses important events in the signal and an error rate more that of energy-aware intermittent sampling without the ASA (figure \ref{fig:battery-based} (a)).

\subsection{Expiration Mitigation via Sample Queue Dynamics}
Instead of enforcing freshness only when samples expire, we propose anticipating and preventing latency violations by analyzing sample queue dynamics. As intermittent systems operate, they enqueue and dequeue their sample queue at various rates based on energy availability and mode of operation. We refer to these rates as \textbf{sample queue dynamics}.

Prior freshness enforcement mechanisms that drop samples do not align well with time buffering---as shown in \ref{fig:interm-adaptive-drops}---because expired samples are opportunity costs. Expired samples both (1) fail to provide data to the server and (2) consume valuable energy that could have been better spent to ensure other sample data does get to the server. We can avoid the opportunity cost by mitigating expirations, ensuring that any energy spent on sampling contributes to the overall accuracy of the data.

We enforce this mitigation by analyzing sample queue dynamics. As the intermittent system with time buffered ASA operates, enqueuing and dequeuing rates vary based on the ASA. Certain enqueuing and dequeuing rates are more likely to fill the sample queue and cause future sample expirations. For example, a period of significant samples can enqueue so many samples that no future processing and communication can dequeue in time; these unsafe rates cause expirations. Hence, if the current queue dynamics match these dangerous rates, \sysname{} limits the ASA to slower rates to avoid sample expirations. 

Time buffering is central to ensuring ASA operation on signal-aware intermittent systems without frequent power failure. But time buffering is not without issues, so analyzing sample queue dynamics to mitigate expiration keeps the time buffering under control and prevents it from becoming disadvantageous. Through this combined operation, \sysname{} enables the accuracy of signal-awareness on intermittent systems. The next section discusses the technical aspects of this design in more depth.

\begin{figure}
    \centering
    \includegraphics[width=0.48\textwidth]{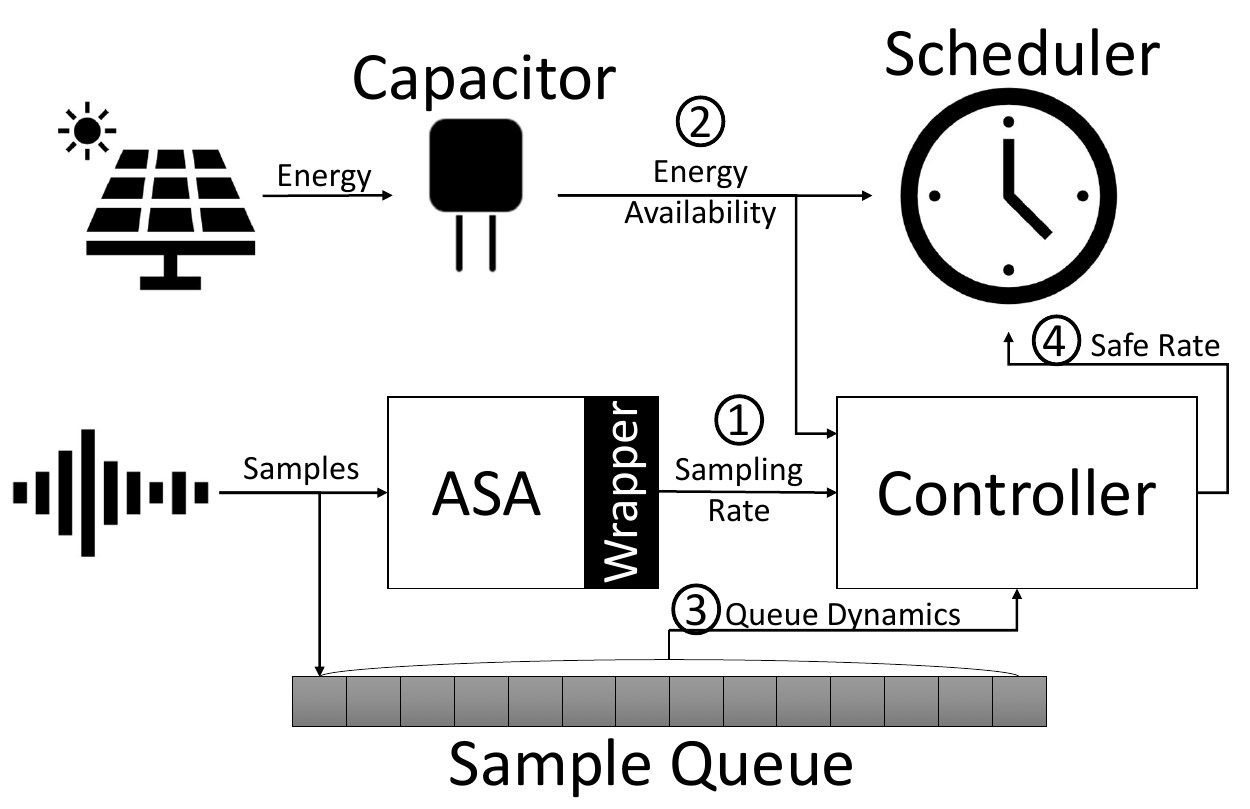}
    \caption{\sysname{}'s overview}
    \label{fig:overview}
\end{figure}

\section{\sysname{} Framework}

\sysname{} requires only small changes to work with existing intermittent frameworks and adaptive sampling algorithms. Thus, for brevity we focus on \sysname{}'s novel aspects: a wrapper for interfacing with ASAs (subsection \ref{subsec:wrapper}), a scheduling policy for time buffering (subsection \ref{subsec:scheduler}), and a controller for analyzing sample queue dynamics and enforcing expiration mitigation (subsection \ref{subsec:controller}). 

Figure \ref{fig:overview} depicts the \sysname{} framework on an intermittent setup. The sampling rate is determined by the ASA (1). The controller considers sampling rate (1) energy availability (2), and sample queue dynamics (3) to determine a safe sampling rate (4), balancing signal-awareness and sample expiration mitigation. Finally, the safe sampling rate is communicated to the scheduler (4), which schedules sampling, processing, and communication based on the safe sampling rate. While these additional mechanisms can add significant overhead to intermittent runtimes, a well-engineered implementation can achieve negligible overhead (Subsection \ref{subsec:overhead}).

\subsection{ASA and the Wrapper}\label{subsec:wrapper}
The ASA module accomplishes the signal-awareness aspect of \sysname{} by determining the current sampling rate based on prior samples. For instance, the linear ASA \cite{ASA_Linear} alters its rate based on differences in consecutive samples. This ASA considers big differences to be significant, so it increases the sampling rate if these differences are more than expected and reduces sampling rate if they are not. While various ASAs with their own definitions of significant exist, the majority work based on the above principle of comparing recent samples to a threshold that determines significance \cite{ASA1, ASA_Linear, ASA_Deviation} To ease communication between system components and different ASAs, we also introduce an ASA wrapper.

\subsection{Scheduler}\label{subsec:scheduler}
The scheduler module manages the time buffering. Prior schedulers determined the sampling rate internally based on energy availability; \sysname{}'s scheduler leaves the sampling rate determination to the controller. This external sampling rate determines when the scheduler should schedule sampling tasks. These tasks consume some ratio of the available harvested energy. The scheduler then uses the rest of the energy to schedule processing and communication tasks. This strategy maintains a constant energy consumption rate for an intermittent system and achieves time buffering through postponing processing and communication tasks.

\begin{table}[htbp]
\caption{Properties of Evaluation Datasets. }
\begin{center}
\begin{tabular}{l c c c c}
\textbf{Dataset}&\textbf{Labels}&\textbf{Mean}&\textbf{Std. Deviation}&\textbf{CV}\\
\hline
EOG\cite{EOG}               &12 & 101.17& 113.81& 1.12\\
Epilepsy\cite{Epilepsy}     &4  &   1.39&   0.69& 0.50\\
Passwords\cite{Passwords}   &5  &   0.81&   0.59& 0.73\\
Strawberry\cite{Strawberry} &2  &   0.72&   0.69& 0.95\\
Tiselac\cite{Tiselac}       &9  & 806.43& 333.84& 0.41\\
Trajectories\cite{Characters} &20 & 1.45&   0.78& 0.54\\
UCI\cite{UCI}               &12 &   1.03&   0.19& 0.19\\
\end{tabular}
\label{tab:dataset}
\end{center}

\end{table}

\subsection{Controller}\label{subsec:controller}
The controller module is in charge of analyzing sample queue dynamics to mitigate expirations. Expirations occur when the system has too many samples in the queue and dequeuing rate is not sufficient to process and communicate the samples by their latency deadlines. We refer to the sample queue count in this case as the \textit{critical} count. 

When the system has a critical number of samples in the queue, the controller reduces the maximum possible ASA rate to a uniform rate which matches the dequeue rate. In doing so, the system still takes significant samples, but does so at a predictable rate. This strategy matches the enqueuing and dequeuing rate, ensuring that the queue size does not increase in the critical phase and thus no expirations can occur (as the oldest sample is always dequeued). However, when taking less significant samples, the system follows an adaptive strategy, which reduces the queue size. This rate reduction remains in place until the system moves out of the critical phase.

\section{Experimental Results}\label{sec:eval}
\subsection{Experimental Setup}
We evaluate \sysname{} with our in house simulator on a variety of configurations.\footnote{All code and datasets will be publicly released after the anonymous reviewing.} The simulation operates as follows. We use sensor datasets from table \ref{tab:dataset} as ground truth for our sensing simulations. These datasets represent a variety of events and exhibit various signal properties. Sampling policies (either uniform or an ASA) determine which entries should be taken from these datasets. The intermittent system collects the samples, processes the collected data and transmits them to a server. At each step, the simulation ensures that sufficient energy is present to accomplish a task, regardless of whether the simulation is in battery-backed or intermittent mode. Failure occurs otherwise, reducing accuracy. Latency violations result in sample expirations, which are not transmitted to the server and thus reduce accuracy. The server reconstructs the full data sequence using linear interpolation to infer missing data points and measures the mean absolute error (MAE) between the reconstructed signal and ground truth.
\begin{figure*}
    \centering
    \includegraphics[width=\textwidth]{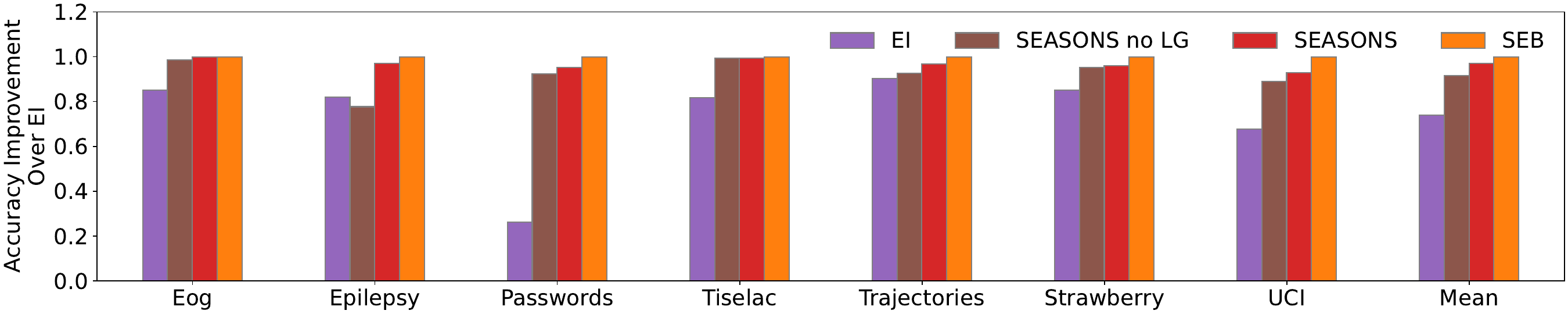}
    \caption{Accuracy improvement of \sysname{} over energy-aware intermittent w/ uniform sampling. \sysname{} improves the accuracy of intermittent systems and approaches the feasible accuracy of a battery-backed ASA.}
    \label{fig:general1}
\end{figure*}
\begin{table}[htbp]
\caption{Example Power Cost of Operation in an Example Edge Sensor.}
\begin{center}
    \begin{tabular}{c|c|c}
    \textbf{Operation}&\textbf{Power Cost (\si{\milli\watt})}&\textbf{Ratio of Total Power}\\
    \hline
    Sampling&2.1&20\%\\
    Encryption&0.2&1\%\\
    BLE&8.4&79\%\\
    Total&10.7&100\%\\
    \end{tabular}
    \label{tab:ecost}
\end{center}
\end{table}

We use the power requirements for sampling, encryption, and communication from table \ref{tab:ecost} over a one second window as guide for the power consumption metrics in our simulation. 
The application samples 50 accelerometer measurements using an IMU \cite{IMU}, encrypts them on a TI MSP430 FR5994 MCU \cite{MSP430} and communicates them via a CC2640R2 BLE chip \cite{CC2640R2}. The power measurements are taken using the EnergyTrace tool \cite{EnergyTrace}. Table \ref{tab:ecost} displays the measured values.

We compare the following, 
\begin{itemize}
    \item \textit{Energy-Aware Intermittent (EI)}. This represents state-of-the-art energy aware intermittent systems that use uniform sampling to meet energy constraints, \textit{e.g.} CatNap \cite{Maeng_CatNap} \& REHASH \cite{Bakar_REHASH}.
    \item \textit{Signal- \& Energy-Aware Battery-backed (SEB)}. This represents a state-of-the-art ASA with a battery \cite{ASA_Linear}.
    \item \textit{\sysname{} no Latency Guarantee (LG)}. \sysname{} without the analysis done by the controller. This configuration does not anticipate and mitigate sample expirations, but will not communicate expired samples). 
    \item \textit{\sysname{}}. The approach in this paper: signal- and energy-aware intermittence while respecting user-defined latency constraints.
\end{itemize}

We experiment with the above configurations over a range of latency and energy constraints to properly evaluate \sysname{}. To compare improvement in accuracy, we measure improvement over EI normalized to SEB.

\subsection{Accuracy Improvements with \sysname{}}
We simulate each configuration on the datasets from table \ref{tab:dataset} with 7 different energy constraints and reasonable latency constraints to understand \sysname{}' accuracy improvements over prior work. For each run, we report the accuracy improvement in figure \ref{fig:general1}. Overall, \sysname{} without latency guarantees (brown), \sysname{} (red), and the battery-backed (orange) configuration achieve respectively 22\%, 31\%, and 35\% improvements in accuracy over the uniform strategy employed by prior intermittent systems (purple).

Compared to the intermittent baseline, \sysname{} without latency guarantees improves accuracy, but does not do so reliably. For instance, with the epilepsy dataset, \sysname{} without latency guarantees results in many sample expirations and produces accuracy below existing intermittent systems. In contrast, \sysname{} with latency guarantees reliably improves accuracy for all datasets, achieving close to existing ASAs, but without a battery. {\it This scenario demonstrates that \sysname{} correctly combines an intermittent power setup with an adaptive sampling algorithm to gain signal-aware accuracy benefits with a batteryless design. }

\begin{figure}
    \centering
    \includegraphics[width=0.48\textwidth]{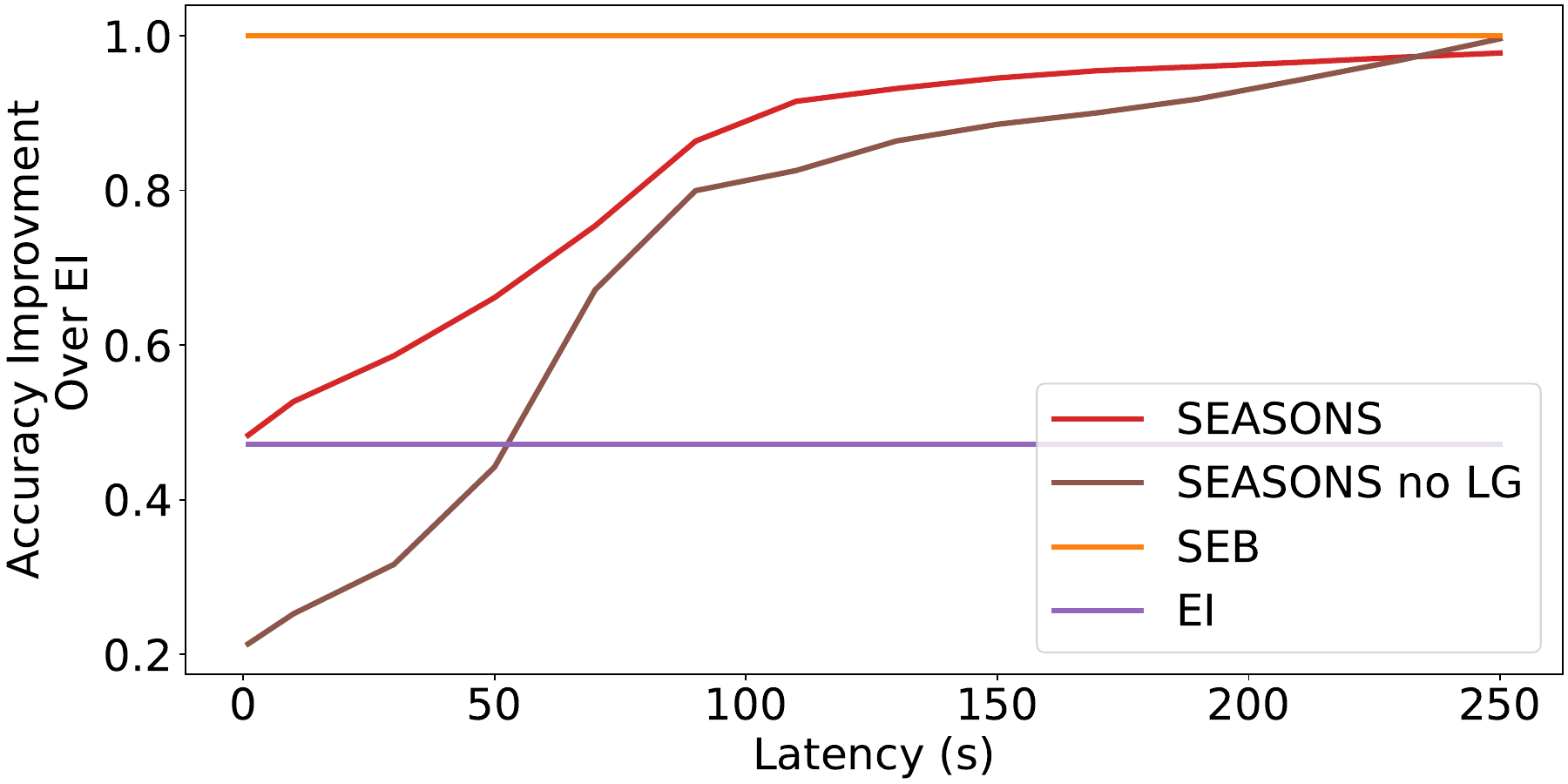}
    \caption{\sysname{} Comparison between EI, SEB, \& EI w/ \sysname{} as latency constraint changes for a collection rate of 60\%.}
    \label{fig:latencyrange}
\end{figure}

\subsection{Latency Sensitivity of \sysname{}}

We simulated each configuration on a range of latency constraints on the \textit{Linear} ASA \cite{ASA_Linear} to investigate the accuracy impact of the user-specified latency. For brevity, we only show accuracy improvements for the UCI dataset in figure \ref{fig:latencyrange} (other datasets exhibit a similar trend). For very short constraints, \sysname{} achieves accuracy similar to that of an energy-aware configuration, which is reasonable given that there is little room for postponability. On the other hand, increasing latency constraints improves accuracy as \sysname{} operates more akin to a battery powered system with an ASA. Moreover, \sysname{} with no latency guarantees follows a similar trend, but performs much worse given tight constraints.

This experiment demonstrates relaxing latency constraints improves \sysname{}' accuracy improvement over prior work; however, \sysname{} improves accuracy even with tight latency constraints. Further, \sysname{}' latency guarantees are vital in ensuring that \sysname{} is beneficial to accuracy, even given short constraints. In summary, {\it \sysname{} is never worse than existing approaches, and given any latency flexibility improves on existing work.}

\subsection{Overhead of \sysname{}}\label{subsec:overhead}
To measure the overhead of \sysname{}, we employ a uniform policy (as in prior intermittent work) and \sysname{} on the TI MSP430FR5994 \cite{MSP430} and emulate their operation on the UCI dataset.
\sysname{} retains the accuracy improvements of the simulation, but incurs an overhead of 1.6x compared to the uniform policy. However, the 0.6x increase in power consumption amounts to only \SI{20}{\micro\watt}, less than 0.2\% of the entire system's power consumption and 2 orders of magnitude less than sampling or communication (table \ref{tab:ecost}). \textit{Hence, \sysname{}'s overhead is negligible.}

\section{Related Work}\label{sec:related}

\paragraph*{Intermittent Systems}
Batteryless deployment and its benefits have been discussed extensively in prior work \cite{Maeng_CatNap, Lucia_Dino, Desai_Camaroptera, Colin_Chain, Gobieski_intelligence, Hester_MayFly,Yildrim_InK, Denby_Orbital, CircusMaximus, Bakar_REHASH}. Frequent checkpointing \cite{Balsamo_Hibernus} and task based programming \cite{Colin_Chain} are popular methods of ensuring forward progress in prior work. CatNap \cite{Maeng_CatNap} \& REHASH \cite{Bakar_REHASH} are state of the art energy-aware intermittent systems that adapt to scarce energy conditions. To the best of our knowledge, no prior intermittent systems have deployed adaptive sampling because they could not support variable energy consumption.

\paragraph*{Adaptive Sampling Algorithms}
ASAs employ signal awareness to balance energy consumption and sensor data accuracy on low power embedded systems. Some algorithms rely on linear or autoregressive models \cite{ASA1, ASA_Linear, ASA_Deviation}, while others employ sophisticated methods such as Reinforcement Learning \cite{ASA_Reinf1} or Neural Networks \cite{ASA_SkipRNN, BudgetRNN_Tejas}. To our knowledge, all prior ASAs assume a large energy buffer and cannot be directly combined with batteryless systems.

\section{Conclusion}
\sysname{} is the first framework to achieve signal- \& energy-aware operation on a batteryless intermittent operation. Through Time Buffering, the ASA employed by \sysname{} operates without requiring a big energy bank (\textit{e.g.} battery). Further, \sysname{} efficiently avoids expirations via analyzing Sample Queue dynamics. The framework operates with minimal overhead and can achieve 31\% more accuracy compared to traditional energy-aware intermittent sensing. Given the negligible cost and significant improvements in accuracy, taking smarter samples in intermittent systems via \sysname{} is worth the effort.

\bibliographystyle{./IEEEtran}
\bibliography{references}

\begin{thebibliography}{10}
\providecommand{\url}[1]{#1}
\csname url@samestyle\endcsname
\providecommand{\newblock}{\relax}
\providecommand{\bibinfo}[2]{#2}
\providecommand{\BIBentrySTDinterwordspacing}{\spaceskip=0pt\relax}
\providecommand{\BIBentryALTinterwordstretchfactor}{4}
\providecommand{\BIBentryALTinterwordspacing}{\spaceskip=\fontdimen2\font plus
\BIBentryALTinterwordstretchfactor\fontdimen3\font minus \fontdimen4\font\relax}
\providecommand{\BIBforeignlanguage}[2]{{%
\expandafter\ifx\csname l@#1\endcsname\relax
\typeout{** WARNING: IEEEtran.bst: No hyphenation pattern has been}%
\typeout{** loaded for the language `#1'. Using the pattern for}%
\typeout{** the default language instead.}%
\else
\language=\csname l@#1\endcsname
\fi
#2}}
\providecommand{\BIBdecl}{\relax}
\BIBdecl

\bibitem{Maeng_CatNap}
K.~Maeng and B.~Lucia, ``Adaptive low-overhead scheduling for periodic and reactive intermittent execution,'' ser. PLDI 2020, 2020.

\bibitem{Lucia_Dino}
B.~Lucia and B.~Ransford, ``A simpler, safer programming and execution model for intermittent systems,'' ser. PLDI '15, 2015.

\bibitem{Desai_Camaroptera}
H.~Desai, M.~Nardello, D.~Brunelli, and B.~Lucia, ``Camaroptera: A long-range image sensor with local inference for remote sensing applications,'' \emph{ACM Trans. Embed. Comput. Syst.}, 2022.

\bibitem{Colin_Chain}
A.~Colin and B.~Lucia, ``Chain: Tasks and channels for reliable intermittent programs,'' ser. OOPSLA 2016, 2016.

\bibitem{Gobieski_intelligence}
G.~Gobieski, B.~Lucia, and N.~Beckmann, ``Intelligence beyond the edge: Inference on intermittent embedded systems,'' ser. ASPLOS '19, 2019.

\bibitem{Hester_MayFly}
J.~Hester, K.~Storer, and J.~Sorber, ``Timely execution on intermittently powered batteryless sensors,'' ser. SenSys '17, 2017.

\bibitem{Yildrim_InK}
K.~S. Y\i{}ld\i{}r\i{}m, A.~Y. Majid, D.~Patoukas, K.~Schaper, P.~Pawelczak, and J.~Hester, ``Ink: Reactive kernel for tiny batteryless sensors,'' ser. SenSys '18, 2018.

\bibitem{Denby_Orbital}
B.~Denby and B.~Lucia, ``Orbital edge computing: Nanosatellite constellations as a new class of computer system,'' ser. ASPLOS '20, 2020.

\bibitem{CircusMaximus}
M.~Afanasov, N.~A. Bhatti, D.~Campagna, G.~Caslini, F.~M. Centonze, K.~Dolui, A.~Maioli, E.~Barone, M.~H. Alizai, J.~H. Siddiqui, and L.~Mottola, ``Battery-less zero-maintenance embedded sensing at the mithr\ae{}um of circus maximus,'' ser. SenSys '20, 2020.

\bibitem{Bakar_REHASH}
A.~Bakar, A.~G. Ross, K.~S. Yildirim, and J.~Hester, ``Rehash: A flexible, developer focused, heuristic adaptation platform for intermittently powered computing,'' \emph{Proc. ACM Interact. Mob. Wearable Ubiquitous Technol.}, 2021.

\bibitem{Balsamo_Hibernus}
D.~Balsamo, A.~S. Weddell, G.~V. Merrett, B.~M. Al-Hashimi, D.~Brunelli, and L.~Benini, ``Hibernus: Sustaining computation during intermittent supply for energy-harvesting systems,'' \emph{IEEE Embedded Systems Letters}, 2015.

\bibitem{ASA_Deviation}
J.~M.~C. Silva, K.~A. Bispo, P.~Carvalho, and S.~R. Lima, ``Litesense: An adaptive sensing scheme for wsns,'' in \emph{2017 IEEE Symposium on Computers and Communications (ISCC)}, 2017.

\bibitem{ASA_Linear}
S.~Chatterjea and P.~Havinga, ``An adaptive and autonomous sensor sampling frequency control scheme for energy-efficient data acquisition in wireless sensor networks,'' in \emph{Distributed Computing in Sensor Systems}, S.~E. Nikoletseas, B.~S. Chlebus, D.~B. Johnson, and B.~Krishnamachari, Eds.\hskip 1em plus 0.5em minus 0.4em\relax Springer Berlin Heidelberg, 2008.

\bibitem{ASA1}
C.~Alippi, G.~Anastasi, C.~Galperti, F.~Mancini, and M.~Roveri, ``Adaptive sampling for energy conservation in wireless sensor networks for snow monitoring applications,'' in \emph{2007 IEEE International Conference on Mobile Adhoc and Sensor Systems}, 2007.

\bibitem{ASA_Reinf1}
A.~Murad, F.~A. Kraemer, K.~Bach, and G.~Taylor, ``Information-driven adaptive sensing based on deep reinforcement learning,'' ser. IoT '20, 2020.

\bibitem{ASA_SkipRNN}
V.~Campos, B.~Jou, X.~Giro-i Nieto, J.~Torres, and S.-F. Chang, ``Skip rnn: Learning to skip state updates in recurrent neural networks,'' 2017.

\bibitem{Subatovich_Ocelot}
M.~Surbatovich, L.~Jia, and B.~Lucia, ``Automatically enforcing fresh and consistent inputs in intermittent systems,'' ser. PLDI 2021, 2021.

\bibitem{ZebraNet}
P.~Zhang, C.~M. Sadler, S.~A. Lyon, and M.~Martonosi, ``Hardware design experiences in zebranet,'' ser. SenSys '04, 2004.

\bibitem{EOG}
F.~Fang and T.~Shinozaki, ``Electrooculography-based continuous eye-writing recognition system for efficient assistive communication systems.'' in \emph{Ambient Assisted Living and Home Care}, 2018.

\bibitem{Epilepsy}
J.~R. Villar, P.~Vergara, M.~Menéndez, E.~de~la Cal, V.~M. González, and J.~Sedano, ``Generalized models for the classification of abnormal movements in daily life and its applicability to epilepsy convulsion recognition,'' 2016.

\bibitem{Passwords}
Graphical password dataset.

\bibitem{Strawberry}
J.~K. Holland, E.~K. Kemsley, and R.~H. Wilson, ``Use of fourier transform infrared spectroscopy and partial least squares regression for the detection of adulteration of strawberry purées,'' \emph{Journal of the Science of Food and Agriculture}, 1998.

\bibitem{Tiselac}
D.~Ienco, R.~Gaetano, C.~Dupaquier, and P.~Maurel, ``Land cover classification via multitemporal spatial data by deep recurrent neural networks,'' \emph{IEEE Geoscience and Remote Sensing Letters}, 2017.

\bibitem{Characters}
B.~H. Williams, M.~Toussaint, and A.~J. Storkey, ``Extracting motion primitives from natural handwriting data,'' in \emph{Artificial Neural Networks -- ICANN 2006}.\hskip 1em plus 0.5em minus 0.4em\relax Springer, 2006, pp. 634--643.

\bibitem{UCI}
D.~Anguita, A.~Ghio, L.~Oneto, X.~Parra, and J.~L. Reyes-Ortiz, ``Human activity recognition on smartphones using a multiclass hardware-friendly support vector machine,'' in \emph{Ambient Assisted Living and Home Care}.\hskip 1em plus 0.5em minus 0.4em\relax Springer, 2012, pp. 216--223.

\bibitem{IMU}
\BIBentryALTinterwordspacing
lsm6dsox datasheet. [Online]. Available: \url{https://www.st.com/resource/en/datasheet/lsm6dsox.pdf}
\BIBentrySTDinterwordspacing

\bibitem{MSP430}
\BIBentryALTinterwordspacing
Ti msp430 fr5994 datasheet. [Online]. Available: \url{https://www.ti.com/lit/ds/symlink/msp430fr5994.pdf}
\BIBentrySTDinterwordspacing

\bibitem{CC2640R2}
\BIBentryALTinterwordspacing
Ti msp430 cc2640r2 datasheet. [Online]. Available: \url{https://www.ti.com/lit/ds/symlink/cc2640r2f.pdf}
\BIBentrySTDinterwordspacing

\bibitem{EnergyTrace}
\BIBentryALTinterwordspacing
Ti msp430 energytrace technology. [Online]. Available: \url{https://www.ti.com/tool/ENERGYTRACE}
\BIBentrySTDinterwordspacing

\bibitem{BudgetRNN_Tejas}
T.~Kannan and H.~Hoffmann, ``Budget rnns: Multi-capacity neural networks to improve in-sensor inference under energy budgets,'' in \emph{2021 IEEE 27th Real-Time and Embedded Technology and Applications Symposium (RTAS)}, 2021.

\end{thebibliography}

\end{document}